\pdfoutput=1
\documentclass[showpacs,superscriptaddress,aps,pra]{revtex4-1}
\usepackage{caption}
\usepackage{subcaption}
\usepackage{cancel}
\usepackage{amsfonts}
\usepackage{amsmath,amsthm}
\usepackage{bm}
\usepackage{amssymb}
\usepackage{mathrsfs}
\usepackage{amsmath, amsthm, amssymb,amsfonts,mathbbol,amstext}
\usepackage{mathtools}
\usepackage[colorlinks=true,citecolor=blue,urlcolor=blue]{hyperref}
\usepackage[normalem]{ulem}
\usepackage{mathrsfs}
\usepackage{dsfont}
\usepackage{graphicx}

\renewcommand{\r}{\rightarrow}
\newcommand{\be}{\begin{equation}}
\newcommand{\ee}{\end{equation}}
\newcommand{\C}{\mathcal}
\newcommand{\tb}{\textbf}
\newcommand{\ti}{\textit}
\newtheorem{theorem}{Theorem}[section]

\newtheorem{definition}{Definition}

\def\be{\begin{equation}}
\def\ee{\end{equation}}

\usepackage{color}
\definecolor{violeta}{cmyk}{0.07,0.90,0,0.34}

\usepackage{cancel}
\usepackage{soul}
\definecolor{cgreen}{RGB}{26, 199, 76}

\begin{document}


\title{Investigating the Global Properties
of a Resource Theory of Contextuality}

\author{Tiago Santos}
\affiliation{Department of Mathematical Physics, Institute of Physics,
University of S\~ao Paulo, R. do Mat\~ao 1371, S\~ao Paulo 05508-090,
SP, Brazil}

\author{Barbara Amaral}
\affiliation{Department of Mathematical Physics, Institute of Physics,
University of S\~ao Paulo, R. do Mat\~ao 1371, S\~ao Paulo 05508-090,
SP, Brazil}

\begin{abstract}
Resource theories constitute a powerful theoretical framework and a tool that captures, in an abstract structure, pragmatic aspects of the most varied theories and processes. For physical theories, while this framework deals directly with questions about the concrete possibilities of carrying out tasks and processes, resource theories also make it possible to recast these already established theories on a new language, providing not only new perspectives on the potential of physical phenomena as valuable resources for technological development, for example, but they also provide insights into the very foundations of these theories. In this work, we will investigate some properties of a resource theory for quantum contextuality, an essential characteristic of quantum phenomena that ensures the impossibility of interpreting the results of quantum measurements as revealing properties that are independent of the set of measurements being made. We will present the resource theory to be studied and investigate certain global properties of this theory using tools and methods that, although already developed and studied by the community in other resource theories, had not yet been used to characterize resource theories of contextuality. In particular, we will use the so called cost and yield monotones, extending the results of reference \cite{wolfe2020quantifying} to  general contextuality scenarios.
    
\end{abstract}

\maketitle

\section{Introduction}
\label{sec:intro}

With the advent of quantum information theory, which brought to physics techniques and methods from computer science, the laws of physics began to be probed through new kinds of questions. In particular, there arose an interest in finding out what is possible within a theory given a set of resources and operations, that is, what the theory allows one to actually perform \cite{horodecki2013quantumness}.  In particular, concepts in foundations of quantum physics began to be investigated in the lights of  a \ti{pragmatic} tradition, in which one is trying to understand and describe how and how much can a physical system be known and controlled through human intervention \cite{coecke2016mathematical}. 

One of the ways in which this pragmatic perspective has come to be formalized by the community is through the so called \emph{resource theories}. A resource theory is a framework that aims for the characterization of physical states and processes in terms of availability, quantification and interconversion of resourceful objects \cite{coecke2016mathematical}. In such a framework, a chosen property is treated as an operational resource \cite{amaral2019resource} and physical phenomena are studied in order to better leverage this specific resource. Two good examples of scientific fields that have  a pragmatic flavour are thermodynamics and chemistry. Both began as endeavors to determine and better understand the ways in which resourceful systems and materials could be transformed and used for one's advantage. Alchemy sought to transform basic metals into nobler ones, and one of the endeavors that marked the early days of thermodynamics was the study of thermal non-equilibrium and its resourcefulness for extracting useful work. Even today, after so much development in both fields, this perspective still drives much of the interest from the community \cite{horodecki2013quantumness}.

Hence, the main concepts behind this kind of approach are resourceful objects and advantageous transformations among these objects. There are many more examples of resource theories and they need not to be extremely practical in purpose or scope. By abstracting the framework one may begin to cast many areas of science in this language and interesting ways of understanding these fields begin to emerge. Even mathematics can be seen as a resource theory in which the resourceful objects are mathematical propositions and the transformations are mathematical proofs, understood as sequences of inference rules \cite{coecke2016mathematical}.

A particularly important class of resource theories are the \ti{quantum resource theories}, resource theories defined in terms of quantum states, processes, protocols and concepts. Quantum resource theories are an example of how to arrive at a particular resource theory from a theory of physics. In it we have a set of processes - state preparations, transformations or measurements, for example - and we divide this set into costly implementable processes and freely implementable ones. Assuming unlimited availability of elements in the free subset, one can then study the \ti{structure} that is induced on the costly set. This kind of resource theory is then  specified by a chosen class of operations, which in the case of a quantum  resource theory is a restriction on the set of all quantum operations that can be implemented. Given this restriction, some quantum states will not be accessible from some fixed initial state and thus become resourceful states which could be harnessed by some agent to reach and end not possible via  the free set only \cite{horodecki2013quantumness}. 

An example of a quantum resource theory is the \ti{resource theory of entanglement}. If we  restrict two or more parties to classical communication and local quantum operations (LOCC), entangled states become resourceful. And thus the full set of quantum states gets separated between the free set of separable states and the costly set of entangled ones. Given access to the free set (separable states), one cannot achieve an entangled state by LOCC. Moreover, access to entangled states allows one to perform tasks such as quantum teleportation that were not possible only via LOCC and the free set of states \cite{horodecki2013quantumness}. There are many more examples of the use of resource theoretic framework in quantum information theory and other areas of physics, such as in the study of asymmetry and quantum reference frames, quantum thermodynamics, quantum coherence and superposition, non-Gaussianity and non-Markovianity \cite{chitambar2019quantum}. Furthermore, it has proven advantageous to recast even more foundational concepts of quantum theory, as contextuality and Bell nonlocality, in resource theoretic frameworks.

Among the advantages of casting a quantum property in resource theory language, we can cite \citep{chitambar2019quantum}:

\begin{itemize}
    \item Resource theories are particularly fitting for restricting our attention to operations and procedures that reflect current experimental capabilities, as generally one can associate a particular resource theory to any specific experiment by taking as free operations only those that can be performed within the limitations of the experimental setup available. Thus, such theory is precisely concerned with the particular tasks that can be done with the setup.
    
    \item Resource theories provide a means of rigorously comparing the \ti{quantity} of resource present in quantum states or channels. As by construction the amount of resource held by an object is at least equal to the amount in another if one can transform the former into the latter by a free operation in the given theory, by studying the interconversion relations in a theory together with the possibilities of quantification, one is able to establish a pre-order on the set of objects within the theory. This ordering structure offers  insight into the role that the property investigated as a resource plays within the bigger theory as a whole. This particular perspective is a great part of this work;
    \item Resource theory allows one to better analyze how and what fundamental processes are responsible for a certain phenomenon. By considering the particular restrictions on the set of operations, one can point out, in a systematic manner, what are the physical requirements for performing a specific task. Interestingly, this can lead one to better consider resource trade-offs through decomposing a certain task in terms of free operations and resource consumption. In certain situations it might be  advantageous to know if by making use of more free objects one can lessen resource consumption.
    
    \item Because the same framework is applicable to  diverse properties, by studying one property of interest within a particular resource theory one can be actually doing much more as it might lead to identification of structures and applications that are common to resource theories in general. As an example we note that ``elegant solutions to the problem of entanglement reversibility emerge when drawing resource-theoretic connections to thermodynamics". 
\end{itemize}
  
In this work, we will investigate some properties of a resource theory for quantum contextuality, an essential characteristic of quantum phenomena that ensures the impossibility of interpreting the results of quantum measurements as revealing properties that are independent of the set of measurements being made \cite{budroni2021quantum}. We will present the resource theory to be studied and investigate certain global properties of this theory using tools and methods that, although already developed and studied by the community in other resource theories, had not yet been used to characterize resource theories of contextuality. In particular, we will use the so called cost and yield monotones, making use of their power in the study of resource theories for non-locality, in an attempt to extend the results of \cite{wolfe2020quantifying} to this more general class of phenomena, contextuality.

This work is organized as follows: in section \ref{sec:resource} we present the basic mathematical elements of a general resource theory; in section \ref{sec:contextuality} we present the resource theory of contextuality considered in this work, defining the set of objects, free objects, and free operations; in section \ref{sec:global} we investigate the global properties of the pre-order of objects defined by the resource theory presented in section \ref{sec:contextuality}; 

\section{Resource theories}
\label{sec:resource}

We begin by describing the basic mathematical elements of a general resource theory \cite{coecke2016mathematical, duarte2018resource, gallego2017nonlocality, amaral2019resource}:

\begin{enumerate}
    \item A set $\C{U}$ of mathematical \ti{objects} that may contain the resource under consideration, together with a subset $\C{F} \subset \C{U}$ whose elements are those which are going to be considered freely available, called \ti{free objects}. 
    
    \item A set $\C{T}$ of transformations between objects that can be freely constructed or implemented, that is, without consuming any resource, called \ti{free transformations}. The notation $A \r B$, in which $A, B \in \C{F}$, denotes that there is a free transformation $F \in \C{T}$ such that $F(A) = B$ and will be used when the specific transformation is not important, but only it's existence. In terms of defining the free transformations, if the free objects are fixed, a transformation $F$ is a free transformation when, for every free object $A$, the resulting object $B = F(A)$ is also a free object.
    
    \item The possibility of combining objects and transformations through binary relations among them. If $A$ and $B$ are objects of the theory, the composite object regarding both is denoted by $A \otimes B$. In a similar manner, if we have two transformations $F$ and $G$, we consider the composite transformation $F \otimes G$ as performing the two transformations in parallel, so that if $F(A)=B$ and $G(C)=D$, then $(F\otimes G) (A \otimes B) = C \otimes D$.
    
\end{enumerate}

Thus we come to a definition of a resource theory, in terms of the elements described above, as follows:

\begin{definition}
    A \tb{resource theory} is defined by the tuple $(\C{U}, \C{F}, \C{T}, \otimes)$, in which $\C{U}$ consists of the set of objects to which the theory refers, $\C{F} \subset \C{U}$ is the set of free objects of the theory, $\C{T}$ is a set of free transformations acting on the objects and a binary operation $\otimes$ that allows parallel combinations of objects and operations.
\end{definition}

For the mathematically oriented reader, we would like to mention that, as \cite{coecke2016mathematical} discusses, this formalization can be summed up by stating that objects and free transformations in a resource theory are, respectively, objects and morphisms in a symmetric monoidal category. In fact, the author of that work also states that ``the difference between a resource theory and a symmetric monoidal category is not a mathematical one, but rather one of interpretational nature'', that ``a particular symmetric monoidal category is called a resource theory whenever one wants to think of its objects as resourceful and its morphisms as transformations or conversions between these resourceful objects''. We refer the reader to references as \cite{coecke2016mathematical, fritz2017resource}.

\subsection{The pre-order of objects}

We now introduce the idea of the \ti{pre-order} of objects in a resource theory. This idea is intimately connected to  interconversion between objects and provides a very natural way of characterizing a given resource theory in terms of an internal structure, the structure of possible interconversions induced by the set of free operations. This idea lies at the heart of this work and we will explore it further.

In a resource theory, sometimes one is not particularly interested in the  process by which a conversion occurs, but rather the important question is whether this conversion is possible or not. That is, given objects $A, B \in \C{U}$, is there a transformation $A \r B$?  

First, since free operations are those that can be done at no cost, it is fairly intuitive that doing nothing is a free operation, that is, for every object $A$, we have $A \r A$. Second, the possibility of freely implementing sequential composition of free operations is also reasonable. By definition, being able of getting from $A$ to $B$ and from $B$ to $C$ at no cost implies being able of getting from $A$ to $C$ at no cost. In other words, we have $A \r B, B\r C \implies A \r C$. These basic facts make of this interconversion relation a \ti{pre-order} in the set of objects, meaning a binary relation that is \ti{reflexive} and \ti{transitive}. Following standard notation we write $A \succeq B$ whenever $A \r B$ in a resource theory, the ``$\succeq$" relation thus defines a pre-order among the objects.

Now, even though this resulting ordered structure is closely related to the specific set $\C{T}$ of free transformations, being actually induced by it, once this set of interconversion relations structure is given, one can ``forget" the transformations that gave rise to the ordering structure and consider only questions about the induced structure itself. In this spirit, one can speak of  \ti{theories of resource convertibility} \cite{coecke2016mathematical}, defined exclusively by the a set of objects equipped with a pre-order and another binary relation:

\begin{definition}
    Given a resource theory $\C{R} = (\C{U}, \C{F}, \C{T}, \otimes)$, the \tb{theory of resource convertibility} associated with $\C{R}$ is the tuple $\tilde{\C{R}} = (\C{U}, \C{F}, \otimes, \succeq)$, in which $\succeq$ is the pre-order relation induced on the objects by the set of free operations.
\end{definition}

 Throughout this work the concept of resource interconvertibility will be  the focus of our discussions,  with one specific choice of free operations for a resource theory of contextuality. We thus feel free to not make further reference to the distinction between a resource theory  and the associated theory of resource interconvertibility.

\begin{definition}
    We redefine a \tb{resource theory} to be the reduced tuple $\C{R} = (\C{U}, \C{F}, \C{T})$ of aforementioned elements together with the induced \tb{theory of resource convertibility} redefined as $\tilde{\C{R}} = (\C{U}, \C{F}, \succeq)$.
\end{definition}

\subsection{Monotones}

One of the most important aspects of a resource theory has to do with  quantifying the \ti{amount of resource} contained in a certain object of the theory. 

\begin{definition}
    Let $(\C{U}, \C{F}, \succeq)$ be a resource theory. We define a \tb{resource monotone} as a function defined on the set of objects, that preserves the the pre-order structure, that is, for all $A, B \in \C{U}$,
\be
M: \C{U} \r \Bar{\mathbb{R}} \text{ \ti{such that}  } B \preceq A \implies M(B) \leq M(A),
\ee
in which $\Bar{\mathbb{R}}$ means the set of extended real numbers $\mathbb{R} \cup \{-\infty,\infty \}$.  

\end{definition} 

Thus a monotone function gives a quantitative measure of the amount of  resource available in an object. Because of their order-preserving property, these functions give us insightful information about the resource theory, as we will see in section \ref{sec:global}.  

It is worth mentioning  that the pre-order structure of objects in a resource theory is  more fundamental  than any single resource monotone. A resource monotone captures certain aspects of the pre-order by assigning numerical values to the objects, but unless the pre-order is a total order (all its elements are comparable), it can never contain the total information available in the pre-order \cite{amaral2019resource}. In fact, even though there were early works in which one of the goals  was to find what would be \ti{the} correct or better resource monotone, the contemporary view is that in general there are several inequivalent monotones and there is no \ti{a priori} reason to choose one over the other. The pre-order is the fundamental structure, with any particular resource monotone being a coarse-grained description of the theory. 

One might  be tempted to question the usefulness of worrying about resource monotones. If they provide only an incomplete description of the total information contained in the pre-order, what does one gain with their use, if anything at all? In general, as it will be in our case, the effort in constructing and investigating resource monotones does pay off and ends up being a crucial part of developing useful resource theories. The authors in  \cite{gonda2019monotones} give an example of the usefulness of resource monotones. They introduce certain properties that they call \ti{global structures} of the pre-order and use resource monotones in the characterization of such properties. Part of this work is exactly trying to answer questions of this nature for a resource theory of contextuality. Another example, is the work \cite{duarte2018concentration}, which shows that contextuality monotones can be used to study geometrical aspects of particular sets of possible behaviors inside and outside the quantum set, as well as \cite{amaral2019resource} and \cite{chitambar2019quantum}, that present and discuss different monotones and their applicability.

\section{A Resource Theory of Contextuality}
\label{sec:contextuality}

Contextuality refers to the impossibility of thinking about statistical results of measurements in quantum systems as revealing pre-existing objective properties of that system,  which are independent of the actual set of measurements one chooses to make  \citep{budroni2021quantum, kochen1975problem}. In this work we deal with the definition of contextuality based in \ti{compatibility scenarios} and the resource theory of contextuality based on \ti{noncontextual wirings}, as defined below.

\subsection{Compatibility scenarios}

\begin{definition}
    Following \cite{amaral2018noncontextual,amaral2019resource, amaral2018graph}, we define a \tb{compatibility scenario} by a triple $\Upsilon := (\C{M},\C{C},\C{O})$, where $\C{M}$ is a finite set of measurements (or random variables) in $(\C{O},\C{P}(\C{O}))$, in which $\C{P}(\C{O})$ is the set of subsets of $\C{O}$, $\C{O}$ is a finite set (representing the  outputs of our measurements),  and $\C{C}$ is a family of subsets of $\C{M}$. The elements $\gamma \in \C{C}$ are the \tb{contexts} of measurements in our scenario.
\end{definition}

Each context $\gamma \in \C{C}$ represents a set of measurements in $\C{M}$ that can be jointly performed. For each context $\gamma$, the set of all possible outcomes for the joint measurement of the measurements in $\gamma$ is the set $\C{O}^\gamma$, that is, each measurement in $\gamma$ can give as result $|\C{O}|$ different outputs. When we jointly perform the measurements of $\gamma$, our output is encoded in a vector $\tb{s} \in \C{O}^\gamma$. 

\begin{definition}
 The \tb{compatibility graph} of a scenario is the graph whose vertices represent the measurements in $\C{M}$ and $x, y \in \C{M}$ are connected by an edge iff there is a context $\gamma \in \C{C}$ such that $x,y \in \C{C}$.
\end{definition}

\subsection{Behaviors}

The main ingredient of our theory, for now, is what we call a behavior.

\begin{definition}
    Given a scenario $(\C{M},\C{C},\C{O})$, a \tb{behavior} $B$ for this scenario is a family of probability distributions over $\C{O}^{\gamma}$, one for each context $\gamma \in \C{C}$,

\be
    B = \Bigg\{ p_\gamma : \C{O}^\gamma \r [0,1]  \Bigg |  \sum_{\tb{s} \in \C{O}^\gamma} p_\gamma (\tb{s}) = 1, \gamma \in \C{C} \Bigg\}
\ee
\end{definition}

To each context $\gamma$ and output $\tb{s} \in \C{O}^\gamma$ the behavior gives the probability $p_\gamma (\tb{s})$ of obtaining  output $\tb{s}$ in a joint measurement of the elements of $\gamma$. Sometimes we also call a behavior a \tb{box}, as a way of creating a mental picture, where we imagine the elements of $\C{M}$ as buttons of the box, and, for each measurement, we imagine the box having $|\C{O}|$ output lights that inform us the result of the measurements.

Behaviors  may or may not satisfy what we call the \ti{non-disturbance condition}, that we states that, given two contexts
$\gamma$ and $\gamma '$, the marginal for their intersection is well defined. If we have, for example, $\gamma = \{x, y\}$ and $\gamma' = \{y, z\}$, the non-disturbance condition implies:

\be
    \sum_{a} p_{x,y} (a,b) = \sum_{c} p_{y,z} (b,c)
\ee
\begin{definition}
 The \tb{non-disturbance set} \tb{ND}$(\Upsilon)$ is the set of behaviors that satisfy the non-disturbance condition for any intersection of contexts in the scenario.
\end{definition}

Another important idea for contextuality is the possibility of assigning a single probability distribution on the whole set $\C{O}^\C{M}$. We call this probability distribution $p_\C{M} : \C{O}^\C{M} \r [0,1]$ a \ti{global section} for the scenario. We say that $p_\C{M}$ is a global section for a scenario $B$ if, in each context, the marginal probability distributions coincide with the ones given by $B$. 

\begin{definition}
 The behaviors that have a global section are called \tb{non-contextual},and the set of non-contextual behaviors will be denoted by \tb{NC}$(\Upsilon)$.
\end{definition}

When a behavior is non-contextual, i.e. when it has a global section, all probabilities can be written as

\be
    p_\gamma (\tb{s}) = \sum_\lambda p(\lambda) \prod_{\gamma_i \in \gamma} p_{\gamma_i} (s_i).
\ee

\subsection{Pre-processing and post-processing operations}

To define the free operations of our resource theory, we begin by defining certain special operations that take behaviors (our objects) in a given scenario into behaviors, possibly in another scenario.

One of the basic operations is the operation of \ti{pre-processing} a behavior. We introduce a new scenario $\Upsilon_{PRE} = (\C{M}_{PRE}, \C{C}_{PRE}, \C{O}_{PRE})$, with new measurements, contexts and outputs, and a new non-contextual behavior $B_{PRE}$ associated with it. We associate each output of $B_{PRE}$ with an input of $B$, in such a way that every output configuration of $B_{PRE}$ defines a possible input configuration in $B$, that is, associated with every output $\tb{r} \in \C{O}_{PRE}$, we have a possible context $\gamma(\tb{r}) \in \C{C}_{PRE}$.

With this, we define a new behavior $\C{W}_{PRE}(B)$ given by

\be
    p_\beta (\tb{s}) = \sum_\tb{r} p_\beta (\tb{r}) p_{\gamma(\tb{r})} (\tb{s}),
\ee
where the sum runs over all outputs $\tb{r}$ associated with the context $\beta$ in $B_{PRE}$.

Analogously, we can define the \ti{post-processing} of a behavior. We again introduce $\Upsilon_{POS} = (\C{M}_{POS}, \C{C}_{POS}, \C{O}_{POS})$ together with a non-contextual behavior $B_{POS}$. The same association is made between outputs $\tb{s} \in \C{O}$ and contexts $\delta (\tb{s}) \in \C{M}_{POS}$. The new behavior obtained $\C{W}_{POS} (B)$ is given by

\be
    p_\gamma (\tb{t}) = \sum_\tb{s} p_\gamma (\tb{s}) p_{\delta (\tb{s})} (\tb{t}).
\ee

\subsection{Non-contextual wirings}

We can now define the free operations we will consider in this work, the  \ti{non-contextual wirings} \cite{amaral2019resource}. We start with an arbitrary behavior $B$ and compose it with a pre-processing $B_{PRE}$ and a pos-processing $B_{POS}$, along with one additional possibility that the probabilities of $B_{POS}$ may also depend on the inputs and outputs of $B_{PRE}$. With this additional freedom, the probabilities of $B_{POS}$ are of the form $p_{\delta} (\tb{t} | \beta, \tb{r})$, but since we want to guarantee that there is no contextuality generated by the processing itself, as done in \cite{amaral2018noncontextual}, we demand that

\be
    p_{\delta} (\tb{t} | \beta, \tb{r}) =  \sum_\phi p(\phi) \prod_{i} p_{\delta_i} (t_i | \beta_i, r_i, \phi).
\ee

With this construction, we get the final scenario $(\C{M}_{PRE}, \C{C}_{PRE}, \C{O}_{POS})$ with an associated behavior $\C{W}_{NC}(B)$, given by

\be
    p_\beta (\tb{t}) = \sum_{\tb{r},\tb{s}} p_\beta (\tb{r}) p_{\gamma(\tb{r})} (\tb{s}) p_{\delta (\tb{s})} (\tb{t})
\ee

This particular class of operations, henceforth referred to as \tb{NCW}, constitutes the free operations of our resource theory. We present here two important results derived in \cite{amaral2019resource} about this class of operations:

\begin{theorem}
    The non-disturbing class of behaviors \tb{ND} is closed under \tb{NCW}.
   \end{theorem}
   
   \begin{theorem}
   The non-contextual class of behaviors \tb{NC} is closed under \tb{NCW}.
\end{theorem}

For the special case of Bell scenarios, a noncontextual wirings is know as a \ti{local operation assisted by shared randomness} (LOSR) \cite{gonda2019monotones}. 

\subsection{Non-contextual deterministic operations}

Now we look for a special class of operations, the set of \ti{deterministic non-contextual wirings}. For that, we  investigate the separability of the wings of the processing operations. For example, writing the pre-processing as

\be
    p_{\beta} (\tb{r}) =  \sum_\phi p(\phi) \prod_{i} p_{\beta_i} (r_i | \phi),
\ee
we see that demanding that $p_{\beta_i}$ takes values in $\{0,1\}$  for each wing $\beta_i$ guarantees that the final form of such a probability distribution is a product of the form

\be
    p^{Det}_\beta (\tb{r}) = \prod_i p^{Det}_{\beta_i} (r_i) = \prod_{i} \delta_{r_i , f_i (\beta_i)}.
\ee

For each wing of the context $\beta$, the function $f_i(\beta_i)$ effectively associates a measurement $\beta_i$ with an output $r_i$, the one for which $p_{\beta_i} (r_i) = 1$. Hence, in a deterministic processing, each context selects a unique output string $\tb{r}$.  This allows us to formulate a definition for a deterministic non-contextual wiring operation. 

\begin{definition}
 We say that a non-contextual wiring operation is \tb{deterministic} when both processings (pre and pos) are deterministic. 
\end{definition}

Then, given a behavior \tb{B}, the behavior $\C{W}^{Det} (B)$ is given by

\be
    p^{Det}_\beta (\tb{t}) = \sum_{\tb{r},\tb{s}} p^{Det}_\beta (\tb{r}) p_{\gamma(\tb{r})} (\tb{s}) p^{Det}_{\delta (\tb{s})} (\tb{t}),
\ee
where the product $p^{Det}_\beta (\tb{r}) p^{Det}_{\delta (\tb{s})} (\tb{t})$ factorizes as

\be
    p^{Det}_\beta (\tb{r}) p^{Det}_{\delta (\tb{s})} (\tb{t}) = \prod_i \delta_{r_i , f_i (\beta_i)} \delta_{t_i , g_i (\delta_i)}
\ee

\begin{definition}
 We define \tb{noncontextual symmetry operations}  as deterministic operations for which the families of functions $\{f_i\}$ and $\{g_i\}$ define one-to-one maps between contexts and outputs.

\end{definition}

For the special case of Bell scenarios, a deterministic non-contextual wiring is know as a \ti{local deterministic operation} (LDO) and a non-contextual symmetry operations is known as  a \ti{local symmetry operation} \cite{gonda2019monotones}.

Just as was  done for \tb{LOSR} operations in \cite{gonda2019monotones}, we can define the \ti{\tb{type}} of a box in terms of the cardinalities of its input and output variables. Since we are dealing with a scenario $\Upsilon = (\C{M}, \C{C}, \C{O})$, the number of input and output variables are fixed by $|\C{C}|$, cardinalities of input variables are fixed by $|\C{M}|$, and cardinalities of output variables are fixed by $|\C{O}|$. Now, as our \tb{NCW} operations take boxes in $\Upsilon = (\C{M}, \C{C}, \C{O})$ to boxes in $\Upsilon_{\C{W}} = (\C{M}_{PRE}, \C{C}_{PRE}, \C{O}_{POS})$, we define the \tb{\ti{type of an operation}} $\C{W}$ as $[\C{W}] \doteq [B] \rightarrow [\C{W} (B)]$. The set of all operations of type $[B_1] \rightarrow [B_2]$ is denoted by $\underset{[B_1] \rightarrow [B_2]}{\text{\tb{NCW}}}$. As mentioned above, some of our results will be type-specific, meaning results concerning type-preserving operations.

\subsection{Convexity of Operations}

An important technical aspect of a resource theory is the convexity of the chosen set of free operations. Convexity is a desirable property  since convex combinations  represent the possibility of choosing  what transformation to implement probabilistically, which should also be considered a free operation.  Let us begin by considering a general free operation  and write it as 

\be
     p_{\beta} (\tb{t}) = \sum_{\tb{r},\tb{s}} p_{\beta, \delta} (\tb{r},\tb{t}) p_{\gamma (\tb{r})} (\tb{s}),
\ee
where we write $p_{\beta,\delta} (\tb{r},\tb{t}) \doteq p_\beta (\tb{r}) p_{\delta (\tb{s})} (\tb{t})$. 

We  ask in which circumstance can we talk about a convex sum of two such operations, say $p^{(0)}_{\beta,\delta} (r,t)$ and $p^{(1)}_{\beta,\delta} (r,t)$. A way of incorporating this notion into the formalism is by making use of the random variables already present in the operations. 

Imagine we want to represent a convex combination where $p^{(0)}_{\beta,\delta} (\tb{r},\tb{t})$ is implemented with probability $\alpha$, and $p^{(1)}_{\beta,\delta} (\tb{r},\tb{t})$ with probability $1 - \alpha$. What we do is to sample from a new binary probability distribution $p (\Lambda)$, in which $p(\Lambda=0) = \alpha$, $p(\Lambda=1) = 1 - \alpha$, such that $\Lambda = 0$ results in $p^{(0)}_{\beta,\delta}$ being implemented, while $\Lambda = 1$ results in $p^{(1)}_{\beta,\delta}$ being implemented. 

Formally, we want to implement 

\be
    \sum_\Lambda p(\Lambda) p_{\beta \r \delta}^{(\Lambda)} (\tb{r} \r \tb{t}) = \sum_\Lambda p(\Lambda) \sum_\lambda p(\lambda | \Lambda) \prod_{i} p_{\beta_i} (r_i | \lambda) p_{\delta_i} (t_i | \beta_i, r_i, \lambda),
\ee
where the extra superscript $(\Lambda)$ in $p_{\beta,\delta}^{(\Lambda)} (r,t)$ denotes the dependence of the operation on the initial sampling over $p(\Lambda)$ through the explicit dependence of $p(\lambda | \Lambda)$ on the variable $\Lambda$. Now, defining $\tilde{p}(\lambda) \doteq \sum_\Lambda p(\lambda | \Lambda) p(\Lambda) $ allows us to write

\be
    \sum_\Lambda p(\Lambda) p_{\beta,\delta}^{(\Lambda)} (\tb{r},\tb{t}) = \sum_\lambda \tilde{p}(\lambda) \prod_i p_{\beta_i} (r_i | \lambda) p_{\delta_i} (t_i | \beta_i, r_i, \lambda),
\ee
which is the standard form of a non-contextual wiring operation. That is, general convex combinations can be naturally incorporated in the formalism, which is surely desirable. When the set of free operations defining a resource theory is convex, we say that the resource theory is a \ti{convex resource theory}. An thus we have just derived an important technical result:

\begin{theorem}
   The set of free operations given by non-contextual wirings is a convex set. Thus, a resource theory of contextuality defined by this set of free operations is a convex resource theory.
\end{theorem}

Now, recalling the discussion on deterministic operations, the convexity of our set of operations gives us another powerful result. Notice that in a general operation with respective contexts $\beta, \delta$, etc, we have that for each particular measurement in each context $\beta_i, \delta_i,...$, we can  write 

\be
p_{\beta_i,\delta_i} (r_i,t_i | \lambda) = \sum_{\Lambda_i} p^{Det(\Lambda_i)}_{\beta_i,\delta_i} (r_i,t_i) p(\Lambda_i | \lambda),
\ee
so that 

\be
\begin{split}
    p_{\beta,\delta} (\tb{r},\tb{t}) & = \sum_\lambda p(\lambda) \prod^{|\beta|}_{i} \sum_{\Lambda_i} p^{Det(\Lambda_i)}_{\beta_i,\delta_i} (r_i,t_i) p(\Lambda_i | \lambda) \\
    & = \sum_{\Lambda_1...\Lambda_{|\beta|}} \prod_i p^{Det(\Lambda_i)}_{\beta_i,\delta_i} (r_i,t_i) \sum_\lambda \prod_j p(\Lambda_j | \lambda) p(\lambda) \\
    & = \sum_{\Vec{\Lambda}} P(\Vec{\Lambda}) \prod_i p^{Det(\Lambda_i)}_{\beta_i,\delta_i} (r_i,t_i),
\end{split}
\ee
in which $\Vec{\Lambda} \doteq (\Lambda_1,...,\Lambda_{|\beta|})$ and $P(\Vec{\Lambda}) \doteq \sum_\lambda \prod_i p(\Lambda_i | \lambda) p(\lambda)$. That is, any operation $p_{\beta,\delta} (r,t)$ is a convex combination of products of deterministic operations, and  we arrive at a generalization of Fine's theorem  \cite{fine1982hidden} and \cite{wolfe2020quantifying}:

\begin{theorem}
    In the resource theory of contextuality defined by non-contextual wirings, the free operations of a given type form a polytope whose vertices are precisely the locally deterministic operations of that type.
\end{theorem}

\section{Investigating the Global Properties
of the Pre-order of Objects}
\label{sec:global}

We finally delve into the main quest of this work: exploring  the possible  interconversion between  objects of our resource theory of contextuality.
Given two behaviors $B_1$ and $B_2$, we say that $B_1$ can be \ti{converted} into $B_2$ if there is a free operation $\C{W} \in$ \tb{NCW} such that $B_2 = \C{W} (B_1)$, in which case we write $B_1 \rightarrow B_2$. If no such operation exists, we write $B_1 \nrightarrow B_2$.

\begin{definition}
In terms of possible relations among two resources, we define:
\begin{itemize}
    \item $B_1$ is \tb{strictly above} $B_2$ when $B_1 \rightarrow B_2$ and $B_2 \nrightarrow B_1$.
    \item $B_1$ is \tb{strictly below} $B_2$ when $B_1 \nrightarrow B_2$ and $B_2 \rightarrow B_1$.
    \item $B_1$ is \tb{equivalent} $B_2$ when $B_1 \rightarrow B_2$ and $B_2 \rightarrow B_1$.
    \item $B_1$ is \tb{incomparable} $B_2$ when $B_1 \nrightarrow B_2$ and $B_2 \nrightarrow B_1$.
\end{itemize}
\end{definition}

\subsection{Cost and Yield monotones}

\begin{definition}
Given any subset of objects $S \subseteq \C{O}$  in a resource theory and a function $f: S \rightarrow \mathbb{R}$ from this set to real numbers, we define the \textbf{yield} and \textbf{cost} relative to $S$ and $f$ as

\be
    Y_{(S|f)} (a) \doteq \max_{\tilde{a} \in S} \{f(\tilde{a}), \mbox{ \ti{such that} } a \rightarrow \tilde{a} \},
\ee

\be
    C_{(S|f)} (a) \doteq \min_{\tilde{a} \in S} \{f(\tilde{a}), \mbox{ \ti{such that} } \tilde{a} \rightarrow a \}.
\ee
Moreover, if there is no such object $\tilde{a}$ such that $a \rightarrow \tilde{a}$ ($\tilde{a} \rightarrow a$), the yield (cost) is set to $-\infty$ ($+\infty$). 
\end{definition}

In other words,  $Y_{(S|f)} (a)$ gives  the value of $f$ for the most resourceful object in $S$ that can be freely obtained from $a$. On the other hand, $C_{(S|f)} (a)$ gives the value of $f$ for the least resourceful object in $S$ from which one can freely obtain $a$.

\subsection{Back to the question of convexity}

One of the goals of a resource theory description is the complete characterization of the pre-order of objects, i.e., actually knowing which of the four possible interconversion relations holds for each pair of objects. The result obtained above about the decomposition of non-contextual free operations into combinations of extremal deterministic ones proves to be actually  useful in this characterization.

Let $\C{P}^{NCW}_{[B_2]} (B_1)$ denote the set of behaviors of type $[B_2]$ that can be obtained by general non-contextual wirings from $B_1$, and $\C{V}^{Det}_{[B_2]} (B_1)$ denote the set of behaviors of type $[B_2]$ obtained from $B_1$  through deterministic non-contextual wirings.
The finite cardinality of the set $\C{V}^{Det}_{[B_2]} (B_1)$ and the existence of a polytope of free operations can be nicely  expressed in the following result:

\begin{theorem}
\be
    \C{P}^{NCW}_{[B_2]} (B_1) = \text{\tb{Conv}}\left(\C{V}^{Det}_{[B_2]} (B_1)\right),
\ee
where $\text{\tb{Conv}}\left(\C{V}^{Det}_{[B_2]} (B_1)\right)$ is the convex hull of the discrete set $\C{V}^{Det}_{[B_2]}(B_1)$.
\end{theorem}

This statement is equivalent to

\be
    B_1 \rightarrow B_2 \iff B_2 \in \text{\tb{Conv}}\left(\C{V}^{Det}_{[B_2]} (B_1)\right),
\ee
which is a very important result, since it actually allows one to check if $B_1 \rightarrow B_2$ through an efficient algorithm. To check if $B_1 \rightarrow B_2$ holds, one has only to determine all the deterministic operations that take behaviors of type $[B_1]$ to behaviors of type $[B_2]$ (which are finite in number), compute the image of $B_1$ under these deterministic operations and then determine whether $B_2$ can be obtained through a convex combination of these images of $B_1$. The answer to last step can be decided with the use of linear programming \citep{wolfe2020quantifying}. 

Now, although the result just obtained is indeed  useful, reducing greatly the number of operations needed to know if $B_1 \rightarrow B_2$, in order to fully characterize the pre-order of objects through this method, one would need to apply the linear program to every pair of objects, which is certainly not practical. Another alternative  to characterize the pre-order is by the use of monotones.

\subsection{Global Properties that characterize a pre-order}
\label{subsec:global}

\begin{definition}
\begin{itemize}
    \item When the pre-order is such that every pair of objects is either strictly ordered or equivalent, the set of objects is said to be \tb{totally pre-ordered}.
    
    \item When there are incomparable objects in the pre-order, if the incomparability relation is transitive among the objects, we say that the pre-order is \tb{weak}.
    
    \item A \tb{chain} is a subset of objects in which \ti{every pair of elements is strictly ordered}. The \tb{height} of the pre-order is the cardinality of the largest chain in this pre-order.
    
    \item Likewise, an \tb{antichain} is a subset of elements in which \ti{every pair of elements is incomparable}. The \tb{width} of the pre-order is the cardinality of the largest antichain contained in the pre-order.
    
    \item We say that an object $B_2$ lies in the \tb{interval} of objects $B_1$ and $B_3$ \tb{i.f.f.} both $B_1 \rightarrow B_2$ and $B_2 \rightarrow B_3$ hold. If the number of equivalence classes in the interval between a pair of objects is \ti{finite} for every pair of inequivalent objects, we say that the pre-order is \tb{locally finite}, otherwise it is said to be \tb{locally infinite}.
\end{itemize}
\end{definition}

Our goal now is the characterization of the resource theory of contextuality defined by non-contextual wirings in terms of these global properties.

\subsection{Monotones: the path to characterize the pre-order}

The monotone construction we will use is partially based on the notion of maximally violating behaviors \cite{amaral2018noncontextual}, behaviors that are analogous to the  PR-boxes  for the CHSH Bell scenario \cite{popescu1994quantum}. Because of this,  we will first  focus on a specific type of contextuality scenario, the $n$-cycle \cite{araujo2013all}, since for this class of behaviors, both non-contextuality inequalities and their quantum violations are known.

The $n$-cycle scenario consists of $n$ dichotomic measurements $X_i$ and a set of contexts of the form $\{X_i, X_{i+1}\}$ modulo $n$. There is a lot to be said about this kind of scenario, but for now let us focus on necessary considerations to define the monotones that will be used henceforth.

\subsubsection*{The $n$-cycle non-contextuality inequalities}

The most general objects we will work with in this scenario are the non-disturbing behaviors. This set of behaviors defines a polytope whose facets are given by the following positivity constrain inequalities \cite{araujo2013all}:

\be
    \begin{cases}
        4p(+ + | X_i X_{i+1}) = 1 + \langle X_i \rangle + \langle X_{i+1} \rangle + \langle X_i X_{i+1} \rangle  \geq 0, \\
        4p(+ - | X_i X_{i+1}) = 1 + \langle X_i \rangle - \langle X_{i+1} \rangle - \langle X_i X_{i+1} \rangle  \geq 0, \\
        4p(- + | X_i X_{i+1}) = 1 - \langle X_i \rangle + \langle X_{i+1} \rangle - \langle X_i X_{i+1} \rangle  \geq 0, \\
        4p(- - | X_i X_{i+1}) = 1 - \langle X_i \rangle - \langle X_{i+1} \rangle + \langle X_i X_{i+1} \rangle  \geq 0,
    \end{cases}
\ee
which are expressed in terms of components of the vector of correlations for simplicity. 

Another important class of inequalities  is the set of inequalities defining the facets of the non-contextual polytope. These are of the form \citep{araujo2013all}:

\be
    \Omega_k = \sum^{n-1}_{i=0} \gamma_i \langle X_i X_{i+1} \rangle \leq n-2,
\ee
with each value of $k$ being associated with a particular choice of values for $\gamma_i \in \{-1, -1\}$ such that the number of $\gamma_i = -1$ is odd. 

A particularly important feature of such inequalities for the constructions and results that will follow is the fact that when we speak of different non-contextuality inequalities in a given $n$-cycle scenario, their respective regions of strict violation are non-intersecting, i.e., for a contextual behavior $B$ there is a unique $k$ for which $\Omega_k (B) > n-2$.

Hence, for a given $n$-cycle scenario (a choice of $n$), one does not have a single non-contextual inequality, but many such inequalities, and each of these inequalities then defines an associated functional $\Omega_k$. If the distinction among the non-contextuality inequalities of a given scenario is unimportant, we shall drop the label $k$ and refer to a general $\Omega$ function for simplicity.

Fortunately, for this specific class of scenarios, not only the form of the inequalities is known but also the value of the associated maximum quantum violations (which are called \ti{Tsirelson bounds}). This problem has interesting connections to graph theory methods, which proved to be fruitful tools for exploring geometric problems of contextuality theory (for discussions of such ideas, see \cite{amaral2017geometrical,amaral2018graph}). The value for the maximal violations are given by \cite{araujo2013all}:

\be
    \Omega_{Q} = \begin{cases} \frac{3n \cos{(\frac{\pi}{n})} -n}{1 + \cos{(\frac{\pi}{n})}} & \mbox{for odd } n, \\ n \cos{(\frac{\pi}{n})} & \mbox{for even } n. \end{cases}
\ee

Behaviors for which the value of the $\Omega$ function is larger than $\Omega_Q$ will be of interest to us, specifically those that \ti{maximally} violate a non-contextuality inequality. This maximal violation can be understood and quantified as follows: since, by construction, we have $|\gamma_i| = 1$, and the correlations obey $|\langle X_i, X_{i+1} \rangle| \leq 1$, it follows that in a general $n$-cycle scenario, the highest value that the function $\Omega$ can have is $\Omega_{\text{PR}} = n$. We call $B_\text{PR}$ the behaviors for which $\Omega(B_\text{PR}) = \Omega_\text{PR}$. We say that such behaviors are \ti{PR-like} by direct analogy with the so called \ti{PR-boxes} defined in the CHSH-scenario as being exactly the behaviors that maximally violate the CHSH inequality\footnote{One may verify that the CHSH-scenario is a particular case of the $n$-cycle scenario, namely the case $n=4$. Accordingly, well established properties of the CHSH scenario can be recovered from the general $n$-cycle properties stated above by setting $n=4$.} \cite{brunner2014bell}.

\subsubsection*{Two useful Cost and Yield monotones}

With these concepts at hand, we are ready to define the functions we will use to characterize the pre-order of objects. 

\begin{definition}[Monotone 1]
We call $M_\Omega$ the yield of a behavior $B$ with respect to the set $\text{\tb{ND}}(\C{N})$ of general non-disturbing behaviors in the $n$-cycle, as measured by the function $\Omega$, that is,

\be
    M_\Omega (B) \doteq Y_{(\text{\tb{ND}}(\C{N}) | \Omega)} (B) = \max_{B^* {\scriptscriptstyle\in} \text{\tb{ND}}(\C{N})} \{\Omega (B^*), \mbox{such that}  B \r B^*\}.
\ee
\end{definition}

We notice that, since the maximization is being carried over the whole set of non-disturbing behaviors, regardless of $B$, any non-contextual behavior can be freely generated after discarding B. In particular, one can always freely choose a behavior $B^*$ with $\Omega(B^*) = n-2$, the highest value a non-contextual behavior can achieve. Hence, the maximum value of $\Omega$ is never less than $n-2$.

To define the second monotone that we are going to use, we need to define some special objects. First, we  define a behavior given by a mixture of a \ti{PR}-like behavior $B_\text{PR}$ and the maximally mixed behavior $B_\varnothing$, which have equal probabilities for all outputs, 

\be
    B = x B_\text{PR} + (1-x) B_\varnothing, \mbox{ {with} } 0 < x < 1,
\ee
such that $\Omega (B) = n-2$, that is, we want this behavior to be in the boundary of the non-contextual set. For this, since $\Omega (B_\text{PR}) = n$ and $\Omega (B_\varnothing) = 0$, we have simply to choose $x = \frac{n-2}{n}$. We call such behavior $B_\textbf{NPR}$ (Noisy-PR), 

\be
    B_\text{NPR} \doteq \frac{(n-2)}{n} B_\text{PR} + \frac{2}{n} B_\varnothing.
\ee

 Finding the suited weight for our convex combination is basically quantifying the amount of noise one has to mix to a \ti{PR}-like behavior for making it non-contextual. As $n$ grows, we notice that the amount of noise necessary gets smaller and smaller, going to zero in the limiting case. Since \ti{PR}-like behaviors lie on the boundary of the non-disturbing polytope and the behavior $B_\text{NPR}$ lies on the boundary of the non-contextual polytope, this gives us  a measure of the volume that the non-contextual set occupies in the full non-disturbing set, in particular how it scales with $n$, \ti{i.e.,} in the limit, the non-contextual set fills in the entire non-disturbing set. Furthermore, knowledge of the quantum, classical and maximal bounds for other scenarios allows one to use this construction to estimate all such relative volumes (classical to quantum, classical to non-disturbing and quantum to non-disturbing) and quantify the scaling of these volumes. The authors in \cite{duarte2018concentration} employ  different techniques to study this kind of geometric characterization of Bell non-locality phenomena and non-local correlations.

Now, returning to our discussion, with this behavior, we define a one-parameter family of mixtures of $B_\text{NPR}$ and $B_\text{PR}$:

\be
    \C{F}_\text{NPR} \doteq \{ F(\alpha): \alpha \in [0, 1] \},
\ee
where $F(\alpha) = \alpha B_\text{PR} + (1 - \alpha) B_\text{NPR}$. Thus, $\alpha$ interpolates between a non-contextual behavior and a maximally violating one (alternatively, $\alpha$ interpolates between the boundaries of the respective polytopes).

Finally, with these tools we can define our second monotone. 

\begin{definition}[Monotone 2] We call $M_\text{NPR}$ the cost of a behavior $B$ with respect to the subset $\C{F}_\text{NPR}$ of non-disturbing behaviors in the $n$-cycle, as measured by the function $\Omega$, that is, 

\be
    M_\text{NPR} (B) \doteq C_{(\C{F}_\text{NPR} | \Omega)} (B) = \min_{B^* {\scriptscriptstyle\in} \C{F}_\text{NPR}} \{ \Omega (B^*), \mbox{ such that } B^* \r B \},
\ee
such that if there is no $B^* \in \C{F}_\text{NPR}$ for which $B^* \r B$, we define this cost to be infinite.
\end{definition}

Now, we have

\be
    \begin{split}
        \Omega (F(\alpha)) &= \alpha \Omega(B_\text{PR}) + (1 - \alpha) \Omega(B_\text{NPR}) \\
        &= \alpha n + (1 - \alpha) (n-2) \\
        & = n + 2(\alpha - 1).
    \end{split}
\ee

With this, $\Omega$ defines a bijection between behaviors on the line $\C{F_\text{NPR}}$ and real numbers. Hence, minimizing $\Omega$ over $B^* \in \C{F_\text{NPR}}$ such that $B^* \r B$ amounts to minimizing the quantity $n + 2(\alpha - 1)$ under the constrain $F(\alpha) \r B$, that is,

\be
    M_\text{NPR} (B) = \min_{\alpha \in [0,1]} \{ n + 2(\alpha - 1), \text{such that} F(\alpha) \r B \}.
\ee

\subsubsection*{Evaluating the monotones}

With these at hand, we now proceed to find closed form expressions for $M_\Omega$ and $M_\text{NPR}$.

Beginning with $M_\Omega$, we already know that $M_\Omega (B) \geq n-2$, where the inequality is saturated by any $B \in \text{\tb{NC}}(\C{N})$, therefore it remains to evaluate the monotone for $B$ non-free. As already mentioned, for a given non-free $B$, there is a unique noncontextuality inequality associated with a functional $\Omega_k$ for which $\Omega_k (B) > n-2$. 

With the fact that every non-free $B$ can be uniquely decomposed into a \ti{PR}-like behavior for $\Omega_k$, with $\Omega_k (B_{\text{PR},k}) = n$, and a free behavior $B_{f, k}$, with $\Omega_k (B_{f, k}) = n-2$, such that $B = \lambda B_{\text{PR},k} + (1- \lambda) B_{f, k}$, we have that such a decomposition gives $\Omega_k (B) = \lambda n + (1- \lambda)(n-2)$. 

Now, consider a general non-contextual operation $\C{W}$, the decomposition above gives 

\be
    \Omega_k \big( \C{W}(B) \big) =  \lambda \Omega_k \big(\C{W}(B_{\text{PR}, k})\big) + (1- \lambda) \Omega_k \big(\C{W}(B_{f,k})\big),
\ee
and since we have $\Omega_k \big( \C{W} (B_{\text{NPR}, k}) \big) \leq n$, and $\Omega_k \big( \C{W} (B_{f, k}) \big) \leq n-2$, it follows that 

\be
    \Omega_k \big( \C{W}(B) \big) \leq \lambda n + (1- \lambda)(n-2) = \Omega_k (B).
\ee

What this means is that for a given non-free $B$ with respect to a function $\Omega_k$, the maximum value of $\Omega_k (B^*)$ for which $B \r B^*$ is the value $\Omega_k (B)$ itself, from which we conclude that

\be
    M_\Omega (B) = \begin{cases}
        n-2 \mbox{,  for } B \mbox{ free,} \\ \Omega_k (B) \text{,  for } B \text{ non-free.}
    \end{cases}
\ee

Turning our attentions to $M_\text{NPR}$, the evaluation is not as straightforward as it was for $M_\Omega$. First, recall the definition of the behaviors $F(\alpha)$. Now, considering the whole set of non-disturbing behaviors $\text{\tb{ND}}(\C{N})$, let us define the set $\text{\tb{B}}_{b}$ of behaviors $B_b$ lying on the boundary of the non-contextual polytope ($\Omega (B_b) = n-2$). Let us also define the smaller set $\text{\tb{B}}_{bb}$ of behaviors $B_{bb}$ that both saturate $\Omega$ and also lie on the boundary of $\text{\tb{ND}}(\C{N})$. We have, by construction $\text{\tb{B}}_{bb} \subseteq \text{\tb{B}}_{b}$.

Beginning with the case of a non-contextual behavior $B$, since it is a free object, carrying the minimization in the definition of $M_\text{NPR}$ amounts to simply looking for the lowest value of $\Omega(B^*)$ for $B^* = F(\alpha)$ for some $\alpha$. The result obtained above actually assures us that this is achieved by the minimum of $n+2(\alpha -1)$, which is $n-2$, for $\alpha = 0$.  

Now, for the case of non-free behaviors, it will be useful to prove the following auxiliary result: for any $B: \Omega (B) \geq n-2$, the minimization to be carried is equivalent to the following ones,

\begin{flalign}
   \min_\alpha & \Big\{ \Omega (F(\alpha)) \mid F(\alpha) \r B \Big\}, \label{eq:MNPR1}\\
    \min_\alpha & \Big\{ \Omega (F(\alpha)) \mid \exists \gamma \geq 0, \exists \tilde{B}_{b} \in \text{\tb{B}}_{b}, \text{ with } B = \gamma \tilde{B}_{b} + (1- \gamma)F(\alpha) \Big\}, \label{eq:MNPR2}\\
    =&\begin{cases}
        \text{if } B \in \C{F}_\text{NPR}: \Omega (B), \text{ else} \\
        \text{if } B \notin \C{F}_\text{NPR}: n + 2(\alpha -1), \text{ with } \alpha, \gamma \geq 0, \text{ and } \tilde{B}_{bb} \in \text{\tb{B}}_{bb} \text{ all } \\ \text{uniquely defining the decomposition } B = \gamma \tilde{B}_{bb} + (1- \gamma)F(\alpha). \label{eq:MNPR3}
    \end{cases}
\end{flalign}

The first of these quantities is explicitly equivalent to the definition of $ M_\text{NPR}$ given before and is taken as the starting point.

For proving the equivalence between $\eqref{eq:MNPR1}$ and $\eqref{eq:MNPR2}$, we deal separately with the case in which $B \in \C{F}_{\text{NPR}}$ and the case $B \notin \C{F}_{\text{NPR}}$. If $B \in \C{F}_{\text{NPR}}$, then $F(\alpha) \r B$ implies that $B$ is lower on the chain $\C{F}_\text{NPR}$, \ti{i.e.}, one can go freely from $F(\alpha)$ to $B$ by mixing $F(\alpha)$ with $B_\text{NPR} \in \text{\tb{B}}_{b}$, giving us eq. $\eqref{eq:MNPR2}$.

If, on the other hand, $B \notin \C{F}_{\text{NPR}}$, we first recall that

\be
    F(\alpha) \r B \iff B \in \text{\tb{Conv}}\left(\C{V}^{Det}_{[B]} \big(F(\alpha)\big)\right). 
\ee

Now, in order to verify that this implies that $B$ can be generated by mixing $F(\alpha)$ with a behavior in $\text{\tb{B}}_{b}$, we use the notion of a \ti{screening-off inequality}. We say that an inequality $f(B) \geq y$ screens-off the set of behaviors of a fixed type that satisfies it if the set of behaviors that saturate the inequality is free. This notion is  useful when it comes to evaluating statements about behavior convertibility. As an example, consider the question of whether $B_2 \r B_1$. If $B_1$ happens to lie in a screened-off region, the statement that $B_1$ is in the convex-hull of images of $B_2$ under deterministic operations of type $[B_1]$ becomes equivalent to  saying that $B_1$ actually lies in a smaller set: the convex-hull of the images of $B_2$ under deterministic operations of type $[B_1]$ \ti{that are properly inside the screened-off region} plus one boundary point, that is

\begin{multline}
    f(B) \geq y \text{ screens-off } [B_1] \implies B_2 \r B_1 \text{\tb{ iff. }} \exists \tilde{B}^b : f(\tilde{B}^b) = y, \text{ and } \\ B_1 \in \text{\tb{Conv}}\left( \tilde{B}^b, \C{V}^{Det}_{[B_1]} \big(B_2)\big) \cap \{ B: f(B) > y \} \right).
\end{multline}

Therefore, in general, knowledge of such a screening-off inequality allows for convertibility statements regarding screened-off points to be decided through sampling from smaller sets.

For our purposes, the equivalence above is exactly what we need in order to derive eq. $\eqref{eq:MNPR2}$. The screening-off inequality to be considered is naturally $\Omega (B) \geq n-2$, with the associated \ti{screening-off region} being the non-free set, with boundary set $\text{\tb{B}}^b$; the convertibility statement under question being $F(\alpha) \r B$; the result just obtained gives us that for any non-free $B$,

\begin{multline}
F(\alpha) \r B \text{\tb{ iff. }} \exists \tilde{B}_{b} \in \text{\tb{B}}_{b}, \\ \text{ with } B \in \text{\tb{Conv}}\left( \tilde{B}^b, \C{V}^{Det}_{[B]} \big(F(\alpha)\big) \cap \{ B': \Omega(B') > n-2 \} \right).
\end{multline}

Now, it turns out that the only image of $F(\alpha)$ under a deterministic operation that lies in the region screened-off by our inequality is $F(\alpha)$ itself \cite{wolfe2020quantifying}, from which we conclude our desired equivalence

\be
    F(\alpha) \r B \text{\tb{ iff. }} \exists \gamma \geq 0, \exists \tilde{B}_{b} \in \text{\tb{B}}_{b}, \text{ with } B = \gamma \tilde{B}_{b} + (1- \gamma)F(\alpha),
\ee
\ti{i.e., } $\eqref{eq:MNPR1}$ and $\eqref{eq:MNPR2}$ are equivalent. 

Next, we notice that what we have in $\eqref{eq:MNPR2}$ is a minimization to be carried under the constraint of $\alpha$ being such that $B = \gamma \tilde{B}^b + (1-\gamma)F(\alpha)$, and that such a problem could in principle be recast as the following constrained optimization to be carried:

\begin{align*}
    \min_{0 \leq \alpha \leq 1} &\Omega \big( F(\alpha) \big), \text{ such that } \tilde{B}^b \doteq \frac{B - (1-\gamma)F(\alpha)}{\gamma}, \\ &\text{ under the constrain that all probabilities of } \tilde{B}^b \text{ are } \\ &\text{ non-negative, with } \gamma \text{ being an implicit function of } \alpha \text{ given by } \\ & \Omega \big( \tilde{B}^b \big) = \frac{\Omega(B) - (1-\gamma)\Omega \big( F(\alpha) \big)}{\gamma} = n-2.
\end{align*}

Now, we use the argument given in section $B.1$ of \cite{wolfe2020quantifying}, which says that \ti{essentially, this is a constrained optimization problem with a linear objective subject to one linear constraint; namely, that the \tb{smallest} conditional probability in $\tilde{B}_b$ is non-negative. For such optimization problems, it is always the case that the objective is maximized when the constraint is not merely satisfied but saturated, that is, the optimal $\alpha$ arises for some unique $\text{\tb{B}}_{b} \ni \tilde{B}_b = \tilde{B}_{bb} \in \text{\tb{B}}_{bb}$ where the smallest conditional probability is precisely zero}. This argument plus some of the facts established in the preceding paragraphs give us the equivalence between $\eqref{eq:MNPR2})$ and $\eqref{eq:MNPR3}$.

From the considerations above, we finally obtain the desired closed-form expression for $M_\text{NPR}$: if $B$ is free, $M_\text{NPR} (B) = n-2$; for a non-free $B$, there is one non-contextuality inequality and associated function $\Omega$ for which $\Omega(B) > n-2$. Within this region, if $B \in \C{F}_\text{NPR}$, then $M_\text{NPR} (B) = \Omega (B)$; if $B \notin \C{F}_\text{NPR}$, we have 

\begin{multline}
    M_\text{NPR} (B) = n+2(\alpha -1), \text{ where $\alpha$ is such that } \\ B = \gamma \tilde{B}_{bb} + (1 - \gamma) F(\alpha), \text{ with } F(\alpha) \in \C{F}_\text{NPR} \text{ and } \tilde{B}_{bb} \in \text{\tb{B}}_{bb}.
\end{multline}

\subsection{Characterizing the pre-order}

Obtaining closed-form expressions for both $M_\Omega$ and $M_\text{NPR}$ allows us to properly characterize the pre-order of objects of our resource theory. For this, we introduce the following construction of two-parameter families of behaviors

\begin{multline}
    \C{B}_{(*)} \doteq \big\{ B(\alpha, \gamma) : \alpha \in [0,1], \gamma \in [0,1] \big\}, \text{ where } B(\alpha, \gamma) \doteq \gamma B_{bb}^* + (1-\gamma) F(\alpha), \\ B_{bb}^* \text{ is a choice of behavior in } \text{\tb{B}}_{bb}, \text{ and } F(\alpha) \in \C{F}_\text{NPR}.
\end{multline}

Each choice of $B^*_{bb} \in \text{\tb{B}}_{bb}$ defines a family, hence the subscript in $\C{B}_{(*)}$. Moreover, each such family is also given by the convex-hull of $B^*_{bb}$ and the chain $\C{F}_\text{NPR}$, that is, $\C{B}_{(*)} = \text{\tb{Conv}} \left( \big\{ B^*_{bb}, B_\text{PR}, B_\text{NPR} \big\} \right)$.

In terms of our monotones, starting with with $M_\text{NPR}$, for general $(\alpha, \gamma)$, we have

\be
    M_\text{NPR} \big( B(\alpha, \gamma) \big) = n+2(\alpha -1).
\ee

For $M_\Omega$, since $\Omega \big( B(\alpha, \gamma) \big) \geq n-2$ for any $(\alpha,\gamma)$, we have 

\be
M_\Omega \big( B(\alpha, \gamma) \big) = \Omega \big( B(\alpha, \gamma) \big), 
\ee
where, recalling that  $\Omega ( B^*_{bb} ) = \Omega ( B_{NPR} ) = n-2$ and $\Omega ( B_\text{PR} ) = n$, we get

\be
    M_\Omega \big( B(\alpha, \gamma) \big) = n + 2\alpha(1-\gamma) -2.
\ee

Turning now to the proper characterization of the pre-order in terms of the properties introduced in subsection \ref{subsec:global}, we can say:

\medskip
\ti{(1)} Without even considering the whole set $\text{\tb{ND}}(\C{N})$, but looking only at the chain $\C{F}_\text{NPR}$, one sees that between any two given behaviors $F(\alpha_1)$ and $F(\alpha_2)$ there are infinite inequivalent objects, for $\alpha$ runs continuously. Furthermore, since the chain is strictly ordered, each pair of inequivalent objects defines a unique equivalence class. We have, then, an infinite number of such equivalence classes of inequivalent objects between any two behaviors, and hence the chain - and consequently the whole set - is \ti{{locally infinite}}. 

\medskip
\ti{(2)} We also demonstrate that there are incomparable resources in the pre-order. For this, consider the following two objects in $\C{B}_{(*)}$: $B_1 \doteq B\left(0,0\right)$ and $B_2 \doteq B\left(1,\frac{1}{2}\right)$. These behaviors are incomparable, as witnessed by our two monotones:  

\be
    \begin{split}
        &M_\text{NPR} (B_1) = n-2 < M_\text{NPR} (B_2) = n, \text{ and } \\ 
        &M_\Omega (B_1) = n-2 > M_\Omega (B_2) = n-3,
    \end{split}
\ee    
which allows us to conclude that \ti{{the pre-order is not totally ordered}}.

\medskip
\ti{(3)} Next, consider the following three behaviors: $B_1 \doteq B\left(0,0\right)$, $B_2 \doteq B\left(\frac{1}{2},\frac{1}{2}\right)$ and $B_3 \doteq B\left(\frac{1}{2},\frac{3}{4}\right)$. By the same reasoning applied in \ti{(2)}, one may verify that we have $B_1 \nleftrightarrow B_2$ and $B_1 \nleftrightarrow B_3$. But since $\frac{1}{2} B^*_{bb} + \frac{1}{2} B_2 = B_3$ and $B^*_{bb}$ is free, we have that $B_2 \r B_3$, what shows that the incomparability relation is not transitive, hence \ti{the pre-order is not weak}.

\medskip
\ti{(4)} Recalling that the height of the pre-order is the cardinality of the largest chain contained therein, since we know that $\C{F}_\text{NPR}$ is a chain with infinite elements, we conclude that \tb{\ti{the height of the pre-order is infinite}}. To investigate the width of the pre-order, consider the set of points $\{ B(x,x) \mid \frac{1}{2} \leq x \leq 1 \}$, the line segment between points $B\left(\frac{1}{2}, \frac{1}{2}\right)$ and $B\left(1,1\right)$. By inspection, we notice that within this region the function $M_\text{NPR}$ is strictly increasing, while $M_\Omega$ is strictly decreasing, \ti{i.e.}, the pair of monotones witness the incomparability of every pair of objects in this line segment. Hence this line segment constitutes an antichain, and since here there is also an infinite number of incomparable points among each other, by the same logic applied to the height of the pre-order, we conclude that {\ti{the width of the pre-order is also infinite}}.

With this, we have the following:

\begin{theorem}
 The pre-order defined by the set of non-contextual wirings in the set of behaviors of any $n$-cycle scenario  is locally infinite, is not totally ordered, and is not weak. Its height and width are both infinite.
\end{theorem}
\subsection{General scenarios}

We now turn our attention to general compatibility scenarios. First, we recall an important result regarding the necessary and sufficient conditions for contextuality. The $n$-cycle  scenarios are, in some sense, the simplest ones exhibiting quantum non-contextual behaviors.  Besides that, the interest in these scenarios 
comes also from the fact that if the compatibility graph does not contain a cycle, all behaviors  are non-contextual. 

\begin{definition}
Given a graph $\mathrm{G}$,  let $S \subset \mathrm{V}$ be any subset of vertices of $\mathrm{G}$. Then the \emph{induced subgraph} $\mathrm{G}\left[S\right]$
 is the graph whose vertex set is $S$ and whose edge set consists of all of the edges in $\mathrm{E}\left(\mathrm{G}\right)$ that  connect any two vertices in $S$.
\end{definition}

\begin{theorem}
There  is a quantum contextual behavior in the scenario $\Upsilon$ if, and only if, the compatibility graph of the scenario
$ \mathcal{G}$ has an $n$-cycle as induced subgraph with 
$n > 3$.
\end{theorem}

  Equivalently, we may say that there is quantum violation of some noncontextuality inequality for the scenario if, and 
  only if, $\mathcal{G}$ has an $n$-cycle as induced subgraph with 
$n > 3$. In this sense, the $n$-cycle scenarios are the simplest ones where it is possible to find quantum violations of 
noncontextuality
inequalities. For a proof of this result, see reference \cite{BM10}.

Given a compatibility scenario exhibiting quantum contextuality, let $C_n$ be a induced $n$-cycle of its compatibility graph. Consider the set of behaviors for which the measurements outside $C_n$ are all independent and uniformly distributed. This subset is in one to one correspondence with the set of behaviors in the $n$-cycle scenario, and hence the examples constructed in the previous section can be used to extend the results obtained in the previous section to general compatibility scenarios.

\begin{theorem}
 The pre-order defined by the set of non-contextual wirings in the set of behaviors of any compatibility scenario exhibiting contextuality is locally infinite, is not totally ordered, and is not weak. Its height and width are both infinite.
\end{theorem}

\section{Discussion}
\label{sec:discussion}

In this work, we investigate the global properties of the resource theory of contextuality defined by choosing the non-contextual wirings as the free operations of the theory. Our work extends the results of reference \cite{wolfe2020quantifying} from Bell non-locality to  contextuality in general compatibility scenarios, explicitly showing how the free operations of this  resource theory of contextuality are  a natural generalization of LOSR operations for Bell non-locality. Thus, extending some results known for a specific scenario in the Bell non-locality framework, namely, the CHSH scenario, we were able to show that not only our set of free operations forms a polytope, but also were able to provide an algorithmic solution to the problem of deciding whether an arbitrary object of the resource theory can be converted into another. We proceeded then to explore the global properties of the pre-order of objects of our resource theory, and since the methods for exploring these properties were also known for this specific case in the resource theory of Bell non-locality, the effort we spent relating both resource theories paid off. We managed to extend without much difficulty the definitions of proper families of useful resource quantifiers, in particular cost and yield monotones, to a whole class of contextuality scenarios in the more general contextuality framework, which we then explored and were able to characterize, leveraging these families of monotones to finally derive the global properties of this resource theory of contextuality. 

Interestingly, all of the investigated global properties ended up behaving in the general contextuality framework exactly as they behaved in the non-locality framework. Just as happened in the resource theory of non-locality given by LOSR operations, in the resource theory of contextuality defined by non-contextual wirings as free operations, the induced pre-order of objects is locally infinite, not totally pre-ordered, is not weak, and both its height and its width are infinite. These results reveal a great structural similarity between both frameworks. At least in terms of what these properties tell us about the nature of the respective resource theories, it seems that a significant part of how both resource theories are constituted as structures is indeed present in both. It is an interesting question whether this similarity persists if one includes  the possibility of not only single-shot operations, but also multi-copy, asymptotic and non-deterministic conversions.

There are many directions of inquiry one could take at this point. One of them has to do with the completeness of the monotones $M_\Omega$ and $M_\text{NPR}$. Since these monotones are not complete within the Bell non-locality framework, they are also not complete in the general contextuality framework, but we can ask what and how many are the monotones needed to completely characterize the set of objects. 

One can also wonder about the role that symmetry operations play in the kinds of structures we investigated. We explored the role of deterministic operations and in particular showed how, for reasons of convexity, they form the backbone of the structure of the set of free transformations. But what about the specific set of symmetry operations? What is the relationship between them and the polytope of free operations? What kind of structure do them induce on the pre-order of objects?

Finally, a last direction that can be pursued by further research is how to incorporate within our framework the possibility of not only single-shot operations, but also multi-copy, asymptotic and non-deterministic conversions, for example, which could be interesting especially  to experimental implementations. For entanglement, it has been shown that considering stochastic operations changes completely the pre-order defined by the resource theory. It is an open problem to check if this is also the case for non-locality and contextuality.

\begin{acknowledgments}
The authors thank Professor Ana Bel\'en Sainz for discussions. This project was supported financially by  Conselho Nacional de Desenvolvimento Cient\'ifico e Tecnol\'ogico (CNPQ) - Regular program, grant number $131630/2019-9$, CNPq, Chamada Universal 2018, grant number $431443/2018-1$, Funda\~ao de Amparo \`a Pesquisa do Estado de S\~ao Paulo, Aux\'ilio \`a Pesquisa - Jovem Pesquisador, grant number 2020/06454-7, and Instituto Serrapilheira, Chamada 2020. 
\end{acknowledgments}
 
\newpage
\renewcommand\bibname{References}
\bibliographystyle{apalike}
\bibliography{biblio.bib}

\end{document}